# The Reactive and Radiation Electromagnetic Energies of Antennas: a New Formulation

Gaobiao Xiao, *Member, IEEE*

***Abstract*—It is required to calculate the stored reactive energy of an antenna in order to evaluate its Q factor. Although it has been investigated for a long time, some issues still need further clarification. The main difficulty involved is that the reactive energy of an antenna tends to become infinitely large when integrating the conventionally defined energy density in the whole space outside a small sphere containing the antenna. The reactive energy is usually made to be bounded by subtracting an additional term associated with the radiation fields. However, there exists no well-accepted accurate definition for this additional term that is valid for all cases. By re-checking the definition of reactive energies, a new formulation is proposed in this paper which can separate the reactive energy and the radiation energy explicitly based on source-potentials. The clearly defined reactive energy is bounded without necessity to subtract that additional term, and the resultant formulae are easy to implement.**

## I. Introduction

THE stored reactive electromagnetic energy of an antenna is a very important quantity. For example, it can be used to evaluate the Q factor of the antenna and predict its bandwidth. The reactive energy has been investigated by many researchers, and the calculation methods proposed so far can be roughly divided into two categories: (1) methods in early stage based on spherical mode expansion technique [1]-[3], and (2) methods based on determining the fields using computational electromagnetic methods or simulators [4]-[7]. In 1948, Chu in his paper [8] discussed the radiation problem associated with electrically small antennas, and derived ladder type equivalent circuits for $TM_{n0}/TE_{n0}$ spherical waves, by which the upper bound on their Q factors can be predicted. The reactive energy only includes those stored in the reactive elements in the equivalent circuit, hence is bounded and can be accurately evaluated. Collin [1] calculated the reactive energies strictly with fields obtained using mode decomposition method [9][10], where the reactive energies of spherical modes and cylindrical modes are obtained by directly integrating the term $(\varepsilon_0/4)\vec{E}\cdot\vec{E}^*$ and $(\mu_0/4)\vec{H}\cdot\vec{H}^*$ in the whole space outside a sphere with a small radius. Since the integration is infinite, the energy density associated with the radiation fields has to be subtracted from the integrand. Fante [2] extended the results of Collin, and McLean re-examined the case of small antennas and calculated the Q factors of $TM_{10}$ and $TE_{10}$ mode [3]. For small antennas, spherical mode expansion solution for reactive energy is a good approximation, and can provide satisfactory upper bound for Q factors. It has been extended to analyzing antennas with larger sizes [11], where the radiation fields by a current distribution are expanded with spherical modes. However, it is not efficient because many modes may need to be taken into account for antennas with large size and complicated structures. Furthermore, the fields inside the sphere enclosing the antenna cannot be addressed accurately. Therefore, it is more natural to use numerical methods to calculate the reactive energies, as has been investigated by many researchers [5][12]-[14]. Basically, a typical numerical procedure to calculate the reactive energies can be carried out in two steps. The first step is to determine all currents involved in the system, such as the induced currents on metal surfaces, the polarization currents in dielectrics, the magnetization currents in magnetic materials, and the currents in lossy media. This can be done using common methods that have been developed in the computational electromagnetic community, such as methods based on electric field integral equations (EFIEs) [15]-[18], volume integral equations (VIEs) [19][20], and methods based on equivalence principles [21]-[24]. The second step is to calculate the radiation electromagnetic fields from these current sources, and then the stored reactive energies by integrating the energy densities in the whole space. However, as pointed out by many researchers, the stored reactive energy obtained in this way is infinitely large if the conventionally defined electric energy density and magnetic energy density are used. Subtracting an additional term of energy density associated with the radiation fields from the integrand has become a common strategy. Unfortunately, this strategy is not a rigorous solution to reactive electromagnetic energy, and at least three drawbacks need to be addressed. (i) The additional energy density is ultimately an ambiguous concept without a

Gaobiao Xiao is with the Key Laboratory of Ministry of Education of Design and Electromagnetic Compatibility of High-Speed Electronic Systems, the Department of Electronic Engineering, Shanghai Jiao Tong University, Shanghai, 200240, China (e-mail: gaobiaoxiao@sjtu.edu.cn).

rigorous definition. It is usually determined based on spherical waves with their wave centers located at the origin, accordingly, their results are coordinate-dependent [4][25]; (ii) It might be acceptable for small antennas, but may become less accurate for large antennas, in which the propagation pattern is quite different from spherical waves, especially in the region near the radiators; (iii) Theoretically, if we divide the electric field into two parts, for example, $\vec{E} = \vec{E}_{rad} + \vec{E}_{react}$, with $\vec{E}_{rad}$ accounting for radiation electric energy and $\vec{E}_{react}$ for the stored reactive electric energy, then the total electric energy density should include three terms, namely, $0.5\varepsilon_0\vec{E}_{rad}\cdot\vec{E}_{rad}$, $\varepsilon_0\vec{E}_{rad}\cdot\vec{E}_{react}$ and $0.5\varepsilon_0\vec{E}_{react}\cdot\vec{E}_{react}$. Obviously, subtracting $0.5\varepsilon_0\vec{E}_{rad}\cdot\vec{E}_{rad}$ from the total energy may not necessarily result in the reactive electric energy because we cannot prove that the two parts of the electric fields are orthogonal to ensure that $\varepsilon_0\vec{E}_{rad}\cdot\vec{E}_{react}$ will vanish. This inference is true also for magnetic energy.

Yaghjian and Best have adopted this method to calculate the Q factor of antennas [4]. In order to give a clear explanation, the core idea of their formulation is revisited, but only the radiation problems in free space is considered so that we need not take into account of the effect of materials and can focus on the key issue. In this case, (A.8) in [4], which is the starting point of Yaghjian-Best formulation, can be simplified as

$$|I_0|^2 X_0' = \lim_{r\to\infty}\left\{\int_V \mathrm{Re}\left(\vec{B}\cdot\vec{H}^* + \vec{D}^*\cdot\vec{E}\right)dV - r^2\,\mathrm{Im}\int_{4\pi}\left(E\times H_{I_0}'^* - E'\times H_{I_0}^*\right)\cdot\hat{r}d\Omega\right\} \quad (1)$$

where $I_0$ is the excitation current at the feeding port of the antenna. The primes stand for derivatives w.r.t $\omega$, and $X_0$ is the input reactance at the feeding port when it is tuned with a series positive inductance or capacitance. $V$ is the spherical domain with radius $r$. Equation (1) is described as reactive theorem by Rhodes in [26] (the case in free space). The term $\int_V \mathrm{Re}\left(\vec{B}\cdot\vec{H}^* + \vec{D}^*\cdot\vec{E}\right)dV$ is related to the conventionally defined electromagnetic energy. In Yaghjian-Best formulation (1) is transformed into

$$|I_0|^2 X_0' = \lim_{r\to\infty}\left\{\int_V \mathrm{Re}\left(\vec{B}\cdot\vec{H}^* + \vec{D}^*\cdot\vec{E}\right)dV - 2\varepsilon r\int_{4\pi}F^2 d\Omega + \frac{2}{\eta}\mathrm{Im}\int_{4\pi}\vec{F}_{I_0}'\cdot\vec{F}^* d\Omega\right\} \quad (2)$$

where $\varepsilon$ and $\eta$ are respectively the permittivity and intrinsic impedance in free space, and $\vec{F}$ is the far electric field. The first two terms in RHS of (2) are considered as the reactive electromagnetic energy in Yaghjian-Best formulation, which is denoted as

$$W_F = \lim_{r\to\infty}\int_V \mathrm{Re}\left(\vec{B}\cdot\vec{H}^* + \vec{D}^*\cdot\vec{E}\right)dV - 2\varepsilon r\int_{4\pi}F^2 d\Omega \quad (3)$$

Vandenbosch [25] proposed a set of formulae for calculating the reactive energies, which are expressed in closed form of integrations with respect to the current densities in the antenna structure. Again consider the free space situation, the equation (27) in [25], which is the starting equation of Vandenbosch formulation, can be simplified as

$$-\frac{1}{2}\int_{V_s}\vec{E}'\cdot\vec{J}^* dV = P_{rad}' + j\left[\lim_{r\to\infty}\int_V \mathrm{Re}\left(\vec{D}^*\cdot\vec{E} + \vec{B}\cdot\vec{H}^*\right)dV + 2W_{rad}\right] \quad (4)$$

where

$$W_{rad} = \lim_{r\to\infty}\mathrm{Im}\frac{1}{4}\oint_S\left(\vec{E}'\times\vec{H}^* - \vec{E}\times\vec{H}'^*\right)\cdot d\vec{S} \quad (5)$$

Recall that $|I_0|^2 X_0 = -\mathrm{Im}\int_{V_s}\vec{E}\cdot\vec{J}^* dV$, and assume $\vec{J}' = 0$, then (4) is the same as (1). Therefore, the theory bases for Vandenbosch formulation and Yaghjian-Best formulation are almost the same, with a minor difference that the postulation of $\vec{J}' = 0$ in Vandenbosch formulation is a little bit stronger than the postulation of $I_0' = 0$ in Yaghjian-Best formulation. The main difference lies in the way to calculate the reactive energies. In Yaghjian-Best formulation, the reactive energies are calculated with $W_F$. In Vandenbosch formulation, $\left(-0.5W_{rad,G}\right)$ is used as the additional term in [25] to replace the term associated with the radiation power. With this modification, the reactive energy can be directly computed with a set of closed-form expressions that are coordinate- independent. Gustafsson and Jonsson evaluated $W_F$ analytically to get [27],

$$W_F = W_{van} + W_{F_2} \quad (6)$$

where $W_{van} = \left(W_{vac}^m + W_{vac}^e + W_{rad,\,G}\right)$ is the total reactive energy in Vandenbosch formulation, and $W_{F_2}$ is a coordinate- dependent term. If the origin shifts within a small sphere containing the antenna, the variation of $W_{F_2}$ is small.

Yaghjian-Best formulation and Vandenbosch formulation have attracted many attentions from researchers [28]-[33]. They have been successfully applied to analysis and optimization of small antennas [34]-[39], and can be applied for highly dispersive lossy media [40][41]. The Vandenbosch formulation has been extended to time domain [42][43]. However, it is reported that the Vandenbosch formulation can produce negative values of stored energy for electrically large structures [38]. Furthermore, the formulation in time domain may give results that are a little bit different from those obtained with the formulation in frequency domain [43].

There are other methods to calculate the reactive energies [44] A comprehensive comparison of them can be found in [32]. However, for those methods that require to evaluate the reactive energies, $W_F$ and $W_{van}$ are the most popular choices. Unfortunately, it lacks rigorous proof to show either $W_{van}$ or $W_F$ is the correct expression for the reactive energy. Therefore, the issue has not yet been solved completely. There is still no unique definition of stored energy that is widely accepted in literatures and valid in all cases [45].

A careful re-examination on this issue reveals that the difficulty involved in reactive energy can be traced back to an old classical problem: for a given time-varying current distribution $\vec{J}(\vec{r},t)$ in domain $V_a$, how to determine the reactive electromagnetic energy stored in the whole free space. Note that for static fields, the electric energy can be computed in different ways. The most commonly used ones are the following two approaches

$$W_e = \int_{V_a} \frac{1}{2}\rho\phi d\vec{r}_1 = \int_{V_\infty} \frac{1}{2}\vec{E}\cdot\vec{D}d\vec{r}_1 \qquad (7)$$

It is easy to check that the second equation in (7) does not hold true for electro-dynamic fields. Only one integral in (7) is correct for expressing the electric energy in time varying fields. Although $\int_{V_\infty} 0.5\vec{E}\cdot\vec{D}d\vec{r}_1$ is commonly chosen to be the proper one, there exists no rigorous proving or experimental validation. Feynman considered that it was so chosen probably because that the simplest choice is usually the best choice [46].

The above discussed difficulty in calculating the reactive electromagnetic energy of antennas may root in this well known paradox in electromagnetics, which has unfortunately not yet been solved satisfactorily. Based on this observation, the paradox and the basic concept of the reactive energy are revisited in this paper, and a new formulation for calculating the reactive energies of radiators is derived. The reasons for selecting source-potential combinations as the appropriate expressions for reactive electromagnetic energies are provided in Section II. The detailed deduction of the new formulation is described in Section III, and validated with a canonical example in Section IV. The expressions for calculating reactive energies and Q factors of antennas are shown in Section V, numerical examples are provided in Section VI, with a brief summary in Section VII.

## II. Alternative Expressions for Reactive Energies

In order to give a clear description on the formulation, the introducing of the energy densities for static fields are re-examined at first and the derivations are described in details even though some of them are quite fundamental.

Consider a static charge $\rho(\vec{r})$ existing in a bounded region $V_a$ in free space. A popular method to obtain the total energy associated with the charge is to assume that all charges are moved piece by piece from infinite to their current positions. According to energy conservation law, it can be deduced that the total electrostatic energy of the whole system is equal to the work done to the charges, which is derived to be

$$W_\rho = \frac{1}{2}\int_{V_a} \rho(\vec{r}_1)\phi(\vec{r}_1)d\vec{r}_1 \qquad (8)$$

where $\phi(\vec{r})$ is the scalar electric potential with its zero reference point located at infinity, and $\vec{E} = -\nabla\phi$ for static fields. Applying Gauss' Law, $\nabla\cdot\vec{D} = \rho$, (8) can be cast into

$$W_\rho = \frac{1}{2}\oint_{S_\infty} \phi\vec{D}\cdot d\vec{S} + \int_{V_\infty} \frac{1}{2}\vec{E}\cdot\vec{D}d\vec{r}_1 \qquad (9)$$

For the sake of simplicity, hereafter, $S_\infty$ and $V_\infty$ are used to denote the spherical surface and the space when $r\to\infty$. Since $\lim_{r\to\infty}(\vec{D}\cdot\hat{r}) \sim O(1/r^2)$, $\lim_{r\to\infty}\phi \sim O(1/r)$, the first term at the RHS of (9) approaches zero, and it seems natural to define the electric energy density as

$$w_e = \frac{1}{2}\vec{E}\cdot\vec{D} \qquad (10)$$

The magnetic energy associated with steady-state current distribution $\vec{J}(\vec{r},t)$, $\vec{r}\in V_a$, can be expressed by

$$W_J = \frac{1}{2}\int_{V_a} \vec{A}(\vec{r}_1)\cdot\vec{J}(\vec{r}_1)d\vec{r}_1 \qquad (11)$$

where $\vec{A}(\vec{r})$ is the vector magnetic potential relating to the magnetic flux density $\vec{B}$ with $\vec{B} = \nabla\times\vec{A}$. The zero reference point of $\vec{A}(\vec{r})$ also locates at infinity. (11) can again be derived based on energy conservation law by considering the process of exciting a loop with its current increases from zero to a certain value $i_0$. Applying Ampere's Law in static case, $\vec{J} = \nabla\times\vec{H}$, (11) can be transformed to

$$W_J = \frac{1}{2}\oint_{S_\infty} \left(\vec{H}\times\vec{A}\right)\cdot d\vec{S} + \int_{V_\infty} \frac{1}{2}\vec{B}\cdot\vec{H} d\vec{r}_1 \qquad (12)$$

Since $\lim_{r\to\infty}\left(\vec{H}\times\vec{A}\right)\cdot\hat{r} \sim O\left(1/r^3\right)$ for magneto-static fields, the surface integration in (12) also vanishes. Hence the conventional magnetic energy density is then defined as

$$w_m = \frac{1}{2}\vec{B}\cdot\vec{H} \qquad (13)$$

However, for time-varying electric fields, $\vec{E} = -\nabla\phi - \partial\vec{A}/\partial t$, the electric energy (8) has to be changed to

$$\tilde{W}_\rho(t) = \frac{1}{2}\int_{V_a} \rho(\vec{r}_1,t)\phi(\vec{r}_1,t)d\vec{r}_1 = \frac{1}{2}\oint_{S_\infty} \phi\vec{D}\cdot\hat{r}dS + \int_{V_\infty} \frac{1}{2}\left(\vec{E}\cdot\vec{D} + \frac{\partial\vec{A}}{\partial t}\cdot\vec{D}\right)d\vec{r}' \quad (14)$$

The upper script '~' is used to indicate time varying variables. The surface integral still vanishes as $\left(\phi\vec{D}\cdot\hat{r}\right) \sim O\left(1/r^3\right)$. We have therefore

$$\frac{1}{2}\int_{V_a} \rho(\vec{r}_1,t)\phi(\vec{r}_1,t)d\vec{r}_1 = \int_{V_\infty} \frac{1}{2}\left(\vec{E}\cdot\vec{D} + \frac{\partial\vec{A}}{\partial t}\cdot\vec{D}\right)d\vec{r}_1 \neq \int_{V_\infty} \frac{1}{2}\vec{E}\cdot\vec{D}d\vec{r}_1 \quad (15)$$

In time varying case, the Ampere's Law must include the displacement current. Inserting $\vec{J} = \nabla\times\vec{H} - \partial\vec{D}/\partial t$ into (11) yields

$$\tilde{W}_J(t) = \frac{1}{2}\int_{V_a} \vec{A}(\vec{r}_1,t)\cdot\vec{J}(\vec{r}_1,t)d\vec{r}_1 = \frac{1}{2}\oint_{S_\infty} \left(\vec{H}\times\vec{A}\right)\cdot d\vec{S} + \int_{V_\infty} \frac{1}{2}\left(\vec{B}\cdot\vec{H} - \vec{A}\cdot\frac{\partial\vec{D}}{\partial t}\right)d\vec{r}_1 \ (16)$$

Since $\left(\vec{H}\times\vec{A}\right)\cdot\hat{r} \sim O\left(1/r^2\right)$, the surface integral in (16) is not zero but a bounded value. The term of the surface integral is the flux passing through $S_\infty$ and can be interpreted as the energy stored beyond the surface $S_\infty$, which is related to the radiation energy. Apparently, it is reasonable to consider the volume integral in the RHS of (16) as the stored reactive energy associated with the current source, which generally does not equal to $\int_{V_\infty} 0.5\left(\vec{B}\cdot\vec{H}\right)d\vec{r}_1$ for time varying fields.

It is quite clear that, for time varying fields, we have to decide which expression is correct for the electromagnetic energies, or even worse, both of them are not the strictly correct expressions. Contrast to the conventional choice, it is proposed in this paper to select the source-potential combinations as the correct expressions for electromagnetic energies in time varying situations, as some other researchers have proposed [47][48]. For the sake of convenience, only electric energy is taken into account in the following discussions to support this choice.

Firstly, the choice of source-potential is at least directly connected to the energy conservation law. Assume that at a discrete time sequence $t_n$, $n$=1,2,…, $N$, a charge distribution of $\rho\left(\vec{r},t_n\right)$ and the corresponding scalar potential $\phi\left(\vec{r},t_n\right)$ remain electrostatic in free space for a short period at each time point. The electric energy associated with the charge distribution can be calculated in the same way as the electrostatic case, i.e., $0.5\int_{V_a} \rho\left(\vec{r}_1,t_n\right)\phi\left(\vec{r}_1,t_n\right)d\vec{r}_1$, which only depends on the charge density and the potential at time $t_n$, no matter whether the charges are moved to their current positions slowly or fast, along straight lines or curves. In this way, it is reasonable to infer that $0.5\int_{V_a} \rho\left(\vec{r}_1,t\right)\phi\left(\vec{r}_1,t\right)d\vec{r}_1$ is valid for arbitrary time *t*, which is bounded for every *t* if the sources are bounded in region $V_a$. With this choice, the corresponding reactive electric energy can be calculated with volume integral in terms of fields and potentials as

$$\tilde{W}_E = \int_{V_\infty} \left(\frac{1}{2}\vec{E}\cdot\vec{D} + \frac{1}{2}\frac{\partial\vec{A}}{\partial t}\cdot\vec{D}\right)d\vec{r}_1 \qquad (17)$$

However, this Gedanken experiment is not applicable for the choice of $\int_{V_\infty} 0.5\vec{E}\cdot\vec{D}d\vec{r}_1$ because it does not directly connect to the work done to charges, but has to be indirectly connected to the energy through the source-potential term.

Secondly, there is no direct and rigorous proof or experimental validation to support that $\int_{V_\infty} 0.5\vec{E}\cdot\vec{D}d\vec{r}_1$ is valid for time varying fields. Seen from (9) and (17), it is more natural to consider $\int_{V_\infty} 0.5\vec{E}\cdot\vec{D}d\vec{r}_1$ as a secondary quantity relating to $0.5\int_{V_a} \rho\left(\vec{r}_1\right)\phi\left(\vec{r}_1\right)d\vec{r}_1$. Moreover, $\int_{V_\infty} 0.5\vec{E}\cdot\vec{D}d\vec{r}_1$ can be taken as a special case of the modified electric energy (17) for static fields with $\partial\vec{A}/\partial t\cdot\vec{D} = 0$, which also to some extent justifies the new choice.

Thirdly, as pointed out previously, for time varying fields, $\int_{V_\infty} 0.5\vec{E}\cdot\vec{D}d\vec{r}_1$ is infinitely large. Consider the following scenario. The electric energy associated with a static charge in a bounded region is finite in the whole space outside a small sphere containing the charge. However, with the conventional definition of the reactive electric energy, it will *abruptly* become infinitely large when the charge begins to vary with time, no matter how slowly it varies. This is quite unnatural, and no method has been found to remedy this issue satisfactorily. Obviously, there is no such difficulty with the source-potential formulation.

Fourthly, the most possible evidence to justify $\left(0.5\vec{E}\cdot\vec{D}\right)$ as the correct expression of electric energy density for time varying fields may come from the Poynting theorem, which in free space can be expressed as

$$\oint_{S_a} \vec{S}\cdot\hat{n}dS + \int_{V_a} \vec{E}\cdot\vec{J}d\vec{r}_1 = -\frac{\partial}{\partial t}\int_{V_a}\left(\frac{1}{2}\vec{D}\cdot\vec{E}+\frac{1}{2}\vec{B}\cdot\vec{H}\right)d\vec{r}_1 \quad (18)$$

where $\vec{S}$ is the Poynting vector, and the problem region $V_a$ is enclosed by surface $S_a$. The Poynting vector is often regarded as the power flux density. The Poynting relation describes the power balance of the system. Its common interpretation states that the sum of the power flowing out of the surface $S_a$ and the power dissipated in the region should equal to the decreasing rate of the total energy in the region. Therefore, it seems quite natural to define the integral at the RHS of (18) as the total electromagnetic energy and the integrand as the energy density. However, the interpretation of Poynting theorem has always been controversial [49][50]. As a matter of fact, Poynting theorem basically describes the balance between the varying rates of the energies in the system instead of the stored energies themselves. There is no doubt that Poynting equation itself is correct because it is derived directly from the Maxwell equation. However, when we trace back to Poynting's work [51], there was no proof to show that $\left(0.5\vec{E}\cdot\vec{D}\right)$ is exactly the correct expression for electric energy density of time varying electric fields.

The above deduction can also be applied to magnetic fields. According to (16), the magnetic energy in the new formulation can be calculated with volume integral in terms of fields and potentials over the whole space as

$$\tilde{W}_M = \int_{V_\infty}\frac{1}{2}\left(\vec{B}\cdot\vec{H}-\vec{A}\cdot\frac{\partial\vec{D}}{\partial t}\right)d\vec{r}_1 \quad (19)$$

For static fields, $\partial\vec{D}/\partial t = 0$, (19) yields the conventional expression for reactive magnetic energy.

Finally, a case-study with the Hertzian dipole provides a very important support to the new formulation. The analytical results of the electric and magnetic energies obtained using the new formulation exactly agree with those by using Chu's equivalent circuit model, without need to introduce any additional terms like that suggested in [3].

## III. New Formulation for Reactive Energies

For the sake of convenience, we denote

$$\tilde{w}_e(\vec{r}_1,t) = \frac{1}{2}\vec{E}\cdot\vec{D}+\frac{1}{2}\frac{\partial\vec{A}}{\partial t}\cdot\vec{D} \quad (20)$$

$$\tilde{w}_m = \frac{1}{2}\vec{B}\cdot\vec{H}-\frac{1}{2}\vec{A}\cdot\frac{\partial\vec{D}}{\partial t} \quad (21)$$

The stored reactive electric energy can be computed by integrating $\tilde{w}_e$ over the whole space,

$$\tilde{W}_E(t) = \int_{V_\infty}\tilde{w}_e(\vec{r}_1,t)d\vec{r}_1 \quad (22)$$

It can be checked from (15) that $\tilde{W}_E(t)=\tilde{W}_\rho(t)$. So the electric energy can also be calculated with integration over the source distribution region.

Similarly, the stored reactive magnetic energy can be computed by integrating $\tilde{w}_m$ over the whole space,

$$\tilde{W}_M(t) = \int_{V_\infty}\tilde{w}_m(\vec{r}_1,t)d\vec{r}_1 \quad (23)$$

Making use of (16), it can also be calculated with integration over the source distribution region subtracting a surface integral, that is,

$$\tilde{W}_M(t) = \frac{1}{2}\int_{V_a}\vec{A}(\vec{r}_1,t)\cdot\vec{J}(\vec{r}_1,t)d\vec{r}_1 - \frac{1}{2}\oint_{S_\infty}\left(\vec{H}\times\vec{A}\right)\cdot d\vec{S} \quad (24)$$

For pulse sources, the surface integral in (24) is zero since the fields will never reach $S_\infty$ during a limited time period. For harmonic waves, it will be shown later that the time average of the surface integral is also zero. Therefore, in most situations we have $\tilde{W}_M(t)=\tilde{W}_J(t)$, without necessity to evaluate the surface integral at $S_\infty$.

For sinusoidal time varying electromagnetic fields with time dependence of $\exp(j\omega t)$, we have

$$\tilde{w}_e = \frac{1}{4}\vec{E}\cdot\vec{D}^* + \frac{1}{4}j\omega\vec{A}\cdot\vec{D}^* \qquad (25)$$

$$\tilde{w}_m = \frac{1}{4}\vec{B}\cdot\vec{H}^* + \frac{1}{4}j\omega\vec{A}\cdot\vec{D}^* \qquad (26)$$

For the sake of simplicity, the same symbols are used for phasors in the expressions. It can be verified that Poynting relation can be written as

$$\nabla\cdot\vec{S} = \nabla\cdot\left(\frac{1}{2}\vec{E}\times\vec{H}^*\right) = 2j\omega\left(\tilde{w}_e - \tilde{w}_m\right) - \frac{1}{2}\vec{E}\cdot\vec{J}^* \qquad (27)$$

where the term $j\omega\vec{A}\cdot\vec{D}^*/4$ cancel out.

The time-averaged stored reactive energies of an antenna in free space can then be calculated with

$$\tilde{W}_E = \frac{1}{4}\mathrm{Re}\int_{V_a}\left(\rho\phi^*\right)d\vec{r}_1 \qquad (28)$$

$$\tilde{W}_M = \mathrm{Re}\left\{\frac{1}{4}\int_{V_a}\vec{A}\cdot\vec{J}^* d\vec{r}_1 - \frac{1}{4}\oint_{S_\infty}\left(\vec{H}\times\vec{A}^*\right)\cdot\hat{r}dS\right\} \qquad (29)$$

The real part of the surface integral in (29) is found to be zero (See Appendix A), so it can be simplified as

$$\tilde{W}_M = \frac{1}{4}\mathrm{Re}\left\{\int_{V_a}\vec{A}\cdot\vec{J}^* d\vec{r}_1\right\} \qquad (30)$$

The radiation power is calculated in terms of far fields,

$$P_{rad} = \mathrm{Re}\left\{\oint_{S_\infty}\frac{1}{2}\left(\vec{E}\times\vec{H}^*\right)\cdot\hat{r}dS\right\} \qquad (31)$$

Making use of the Poynting theorem in free space, it can also be calculated in terms of the current source and electric field as

$$P_{rad} = \mathrm{Re}\left\{-\int_{V_a}\frac{1}{2}\vec{E}\cdot\vec{J}^* d\vec{r}_1\right\} \qquad (32)$$

With the new formulation for stored reactive energies, the Q factor of a radiator is defined by

$$Q_{po} = \frac{\omega\left(\tilde{W}_E + \tilde{W}_M\right)}{P_{rad}} \qquad (33)$$

Similar to the method proposed by Vandenbosch, (33) can be evaluated at a single frequency alone.

Yaghjian and Best have proposed in [4] two Q factors, namely, $Q_z$ and $Q_{FBW}$. Both of them are defined with the antenna being tuned at $\omega_0$ with a positive series inductance or capacitance. $Q_z$ is derived from the derivative of an input impedance of the antenna,

$$Q_Z = \frac{\omega_0}{2}\frac{\left|R'(\omega_0) + jX_0'(\omega_0)\right|}{R(\omega_0)} \qquad (34)$$

Although requiring the derivatives, it is significant that (34) has avoided evaluating the stored energy. $Q_{FBW}$ is calculated with the fractional bandwidth,

$$Q_{FBW} = \frac{1}{FBW_s(\omega_0)}\frac{s-1}{\sqrt{s}} \qquad (35)$$

where $s$ is the VSWR used to define the fractional bandwidth $FBW_s$ at the tuned angular frequency $\omega_0$. The bandwidth for a given VSWR at a tuned $\omega_0$ can be determined by searching to both sides around $\omega_0$ the reflection coefficient based on the input impedance. $Q_{FBW}$ depends on the choice of VSWR and it can be seen that $Q_{FBW} \approx Q_Z$ when $s \to 1$.

## IV. Validation with Hertzian Dipole

In order to compare the proposed formulation with conventional methods, the stored energy and Q factor of a Hertzian dipole is analyzed, which has been calculated by Mclean in [3]. It's known that a Hertzian dipole generates $TM_{10}$ spherical mode in free space (denoted with $TM_{01}$ in [3]). The fields of a Hertzian dipole are symmetrical about the z-axis, the components of which can be readily derived as below

$$H_\phi = -\sin\theta e^{-jkr}\left(\frac{1}{r} - \frac{j}{kr^2}\right) \qquad (36)$$

$$E_\theta = -\eta \sin\theta e^{-jkr}\left(\frac{1}{r}-\frac{j}{kr^2}-\frac{1}{k^2r^3}\right) \quad (37)$$

$$E_r = 2\eta\cos\theta e^{-jkr}\left(\frac{j}{kr^2}+\frac{1}{k^2r^3}\right) \quad (38)$$

The amplitude of the Hertzian dipole is assumed to be $(j4\pi/k)$ so as that these expressions are exactly the same as those shown in [3]. However, the vector potential given in [3] is not appropriate because it is difficult to find a scalar potential to satisfy the Lorentz Gauge. A proper vector potential can be obtained directly from the dipole current as follows,

$$\vec{A} = \frac{j\mu}{kr}e^{-jkr}\left(\hat{r}\cos\theta - \hat{\theta}\sin\theta\right) \quad (39)$$

The terms relating to reactive energy densities are

$$w_e = \mathrm{Re}\left(\frac{1}{4}\vec{E}\cdot\vec{D}^*\right) = \frac{1}{4}\mu\left[\sin^2\theta\left(\frac{1}{r^2}-\frac{1}{k^2r^4}+\frac{1}{k^4r^6}\right)+4\cos^2\theta\left(\frac{1}{k^2r^4}+\frac{1}{k^4r^6}\right)\right] \quad (40)$$

$$w_m = \mathrm{Re}\left(\frac{1}{4}\vec{B}\cdot\vec{H}^*\right) = \frac{1}{4}\mu\sin^2\theta\left(\frac{1}{r^2}+\frac{1}{k^2r^4}\right) \quad (41)$$

$$-\mathrm{Im}\left(\frac{1}{4}\omega\vec{A}\cdot\vec{D}^*\right) = -\frac{1}{2}\mu\cos^2\theta\frac{1}{k^2r^4}-\frac{1}{4}\mu\sin^2\theta\left(\frac{1}{r^2}-\frac{1}{k^2r^4}\right) \quad (42)$$

As can be checked straightforwardly that the integration of $w_e$ and $w_m$ in the space outside a small sphere with radius $a$ is infinite due to the contribution of the $(1/r^2)$ terms. These terms are canceled in [3] by subtracting the energy density associated with the radiation fields, which is

$$w_{rad} = \frac{1}{4}\varepsilon\left|E_\theta^{rad}\right|^2 = \frac{\mu}{4r^2}\sin^2\theta \quad (43)$$

With the new definition, the integrand of the stored reactive electric energy and magnetic energy are derived to be

$$\tilde{w}_e = \frac{\mu}{4}\left[\sin^2\theta\frac{1}{k^4r^6}+4\cos^2\theta\left(\frac{1}{2k^2r^4}+\frac{1}{k^4r^6}\right)\right] \quad (44)$$

$$\tilde{w}_m = \frac{\mu}{2}\left(\sin^2\theta - \cos^2\theta\right)\frac{1}{k^2r^4} \quad (45)$$

Obviously, there is no $(1/r^2)$ term. Integrating them in the space outside a sphere with radius a, the total reactive energies can be obtained

$$\tilde{W}_E = \frac{2\pi\mu}{3k}\left(\frac{1}{ka}+\frac{1}{k^3a^3}\right),\quad \tilde{W}_M = \frac{2\pi\mu}{3k}\frac{1}{ka} \quad (46)$$

The radiation power is found to be

$$P_{rad} = \frac{4\pi\eta}{3} \quad (47)$$

Hence, the Q factor for the $TM_{10}$ mode is

$$Q_{po} = \frac{\omega\left(\tilde{W}_E+\tilde{W}_M\right)}{P_{rad}} = \frac{1}{ka}+\frac{1}{2k^3a^3} \quad (48)$$

The Q factor is exactly the same as that obtained by Mclean [3]. It can be further verified that, by multiplying a proper scale factor, the reactive electric energy and the reactive magnetic energy are also exactly the same as those stored in the capacitor and the inductor in the equivalent circuit proposed by Chu [8]. Note that the propagating wave of this point source behaves much like a spherical wave. Therefore, subtracting the energy density associated with the radiation power happens to give good results for the total reactive energy in this case.

## V. Expressions for Q Factor of Antennas

Consider a typical radiation problem shown in Fig. 1(a). An excitation current $\vec{J}_{ex}$ exists on the antenna port $S_p$. There is a dielectric with permittivity $\varepsilon_1$ and permeability $\mu$ in region $V_d$, together with a PEC conductor with surface $S_c$.

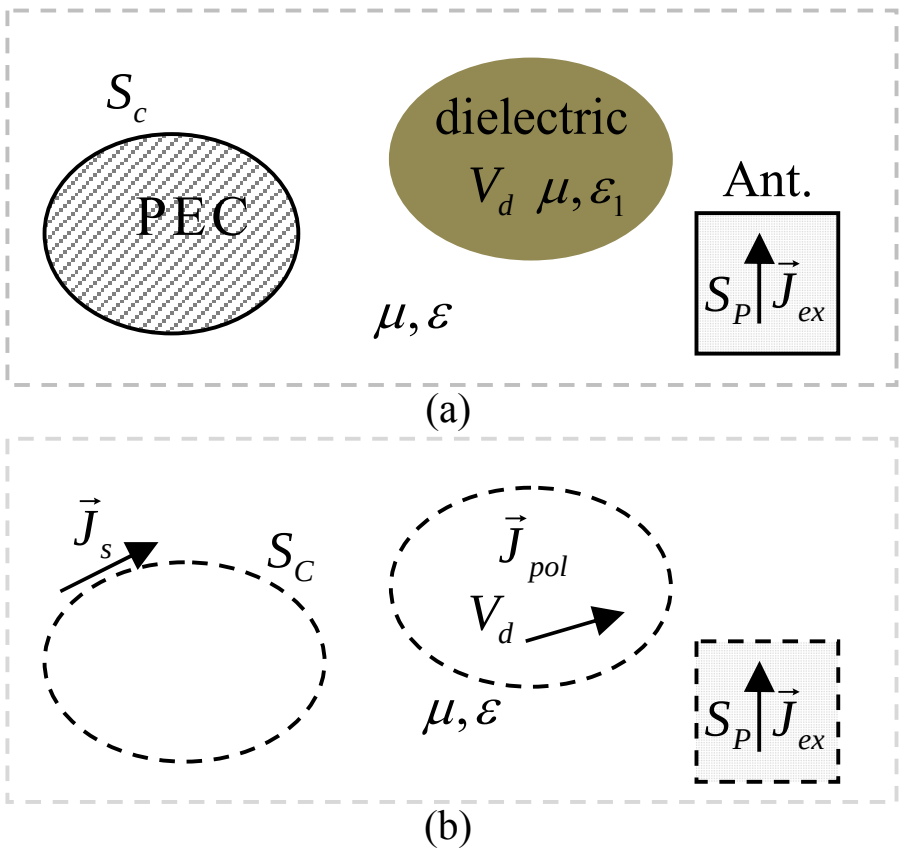

Fig.1. Antenna scattering problem. (a) Antenna near a PEC conductor and a dielectric obstacle; (b) Equivalent problem with all equivalence currents located in free space.

The radiation problem can be solved based on equivalence principle. Assume that there is an induced surface current $\vec{J}_s$ on $S_C$, a polarization current $\vec{J}_{pol}$ in region $V_d$. These equivalent sources are then all placed in free space and are used to account for the effect of the conductor and the dielectric, as shown in Fig.1 (b).

Denote the electric field generated by the excitation current as input field to the conductor and the dielectric, which is denoted by

$$\vec{E}^{in} = \mathcal{L}\left\{\vec{J}_{ex}(\vec{r}_1);\vec{r}_1 \in S_P\right\} \quad (49)$$

The operator $\mathcal{L}\left\{\vec{X}(\vec{r}_1);\vec{r}_1 \in \Omega\right\}$ is defined as

$$\mathcal{L}\left\{\vec{X}(\vec{r}_1);\vec{r}_1 \in \Omega\right\} = -j\omega\mu\left[\int_\Omega g\vec{X}(\vec{r}_1)d\vec{r}_1 + \frac{\nabla}{k^2}\int_\Omega g\nabla_1 \cdot \vec{X}(\vec{r}_1)d\vec{r}_1\right] \quad (50)$$

where $g(\vec{r},\vec{r}_1) = e^{-jkR}/(4\pi R)$ is the scalar Green's function and k is the wave number in free space. $R = |\vec{r}-\vec{r}_1|$. The tangential component of the electric field vanishes on the PEC surface, so we have the electric field integral equation,

$$\left[\mathcal{L}\left\{\vec{J}_c(\vec{r}_1);\vec{r}_1 \in S_c\right\} + \mathcal{L}\left\{\vec{J}_{pol}(\vec{r}_1);\vec{r}_1 \in V_d\right\} + \vec{E}^{in}\right]_{\tan} = 0 \quad (51)$$

In the dielectric, the total electric field includes two parts,

$$\mathcal{L}\left\{\vec{J}_S(\vec{r}_1);\vec{r}_1 \in S_c\right\} + \mathcal{L}\left\{\vec{J}_{pol}(\vec{r}_1);\vec{r}_1 \in V_d\right\} + \vec{E}^{in} = \vec{E} \quad (52)$$

where the polarization current relates to the total electric field ,

$$\vec{J}_{pol} = j\omega(\varepsilon_1 - \varepsilon)\vec{E} \quad (53)$$

Inserting (53) into (52) yields a volume integral equation with respect to the polarization current.

Given an excitation current $\vec{J}_{ex}$, the current $\vec{J}_S$ and $\vec{J}_{pol}$ can be obtained by solving the electric field integral equation of (51) and (52) with method of moment (MoM). The stored energies are then computed with (See Appendix B)

$$\tilde{W}_E = \frac{\mu}{16\pi k^2}\int_{V_\Sigma}\int_{V_\Sigma} \nabla_1 \cdot \vec{J}_\Sigma(\vec{r}_1)\nabla_2 \cdot \vec{J}_\Sigma^*(\vec{r}_2)\frac{\cos kr_{12}}{r_{12}}d\vec{r}_2 d\vec{r}_1 \quad (54)$$

$$\tilde{W}_M = \frac{\mu}{16\pi}\int_{V_\Sigma}\int_{V_\Sigma} \vec{J}_\Sigma(\vec{r}_1)\cdot\vec{J}_\Sigma^*(\vec{r}_2)\frac{\cos kR}{R}d\vec{r}_2 d\vec{r}_1 \quad (55)$$

where $r_{12} = |\vec{r}_1 - \vec{r}_2|$. The integration region is the combination of all source domains, which is denoted by $V_\Sigma = V_d \cup S_C \cup S_P$. The total current is denoted by $\vec{J}_\Sigma = \vec{J}_{pol} \cup \vec{J}_S \cup \vec{J}_{ex}$. Note that $\tilde{W}_E$ and $\tilde{W}_M$ are respectively only the first part of $\tilde{W}_{vac}^e$ and $\tilde{W}_{vac}^m$ given in Vandenbosch formulation.

The Poynting theorem in this case can be expressed as

$$\nabla \cdot \vec{S} = 2j\omega(\tilde{w}_e - \tilde{w}_m) - \frac{1}{2}\vec{E}\cdot\vec{J}_\Sigma^* \quad (56)$$

where the electric field includes contributions from all sources,

$$\vec{E}(\vec{r}) = \mathcal{L}\left\{\vec{J}_S(\vec{r}_1);\vec{r}_1 \in S_C\right\} + \mathcal{L}\left\{\vec{J}_{pol}(\vec{r}_1);\vec{r}_1 \in V_d\right\} + \mathcal{L}\left\{\vec{J}_{ex}(\vec{r}_1);\vec{r}_1 \in S_P\right\} \quad (57)$$

Denote

$$P_{in} = -\frac{1}{2}\mathrm{Re}\left\{\int_{S_P} \vec{E}\cdot\vec{J}_{ex}^{*} d\vec{r}_1\right\}$$
$$P_{lc} = \frac{1}{2}\mathrm{Re}\left\{\int_{S_C} \vec{E}\cdot\vec{J}_{S}^{*} d\vec{r}_1\right\} \quad (58)$$
$$P_{ld} = \frac{1}{2}\mathrm{Re}\left\{\int_{V_d} \vec{E}\cdot\vec{J}_{pol}^{*} d\vec{r}_1\right\}$$

From (56), we have the power balance equation,

$$P_{in} = P_{rad} + P_{lc} + P_{ld} \quad (59)$$

Therefore, the Q factor of the antenna can be denoted as

$$\frac{1}{Q_{in}} = \frac{1}{Q_{rad}} + \frac{1}{Q_c} + \frac{1}{Q_d} \quad (60)$$

All Q factors are defined in the form of (33), with the radiation power therein being replaced by $P_{in}$, $P_{rad}$, $P_{lc}$ and $P_{ld}$, respectively.

For PEC conductors, $P_{lc} = 0$ since the tangential component of the total electric field on the surface is zero. The current on the PEC surface contributes to the storage of reactive energies, but does not contribute to the radiation power directly. However, the surface current will generate electric field on the antenna, hence, indirectly affect the radiation power.

For dielectrics, $\vec{J}_{pol} = j\omega(\varepsilon_1 - \varepsilon)\vec{E} = j\omega(\varepsilon_1' - \varepsilon)\vec{E} - \omega\varepsilon_1''\vec{E}$. The power loss is $P_{ld} = -0.5\omega\varepsilon_1''\vec{E}\cdot\vec{E}^*$. The imaginary part of $0.5\left(\vec{E}\cdot\vec{J}_{pol}^*\right)$ represents reactive energy stored in the dielectric, which has been already addressed in the reactive energies related to the corresponding polarization current. The polarization currents also influence the radiation power by affecting the electric field distribution at the antenna port.

## VI. Numerical Examples

Six examples are analyzed to show the difference between the new formulation and the method proposed in [4] and [25]. In order to get rid of possible ambiguity, the feeding structures are clearly illustrated in the examples. All feeding currents have amplitude of $I_0 = 1\text{A}$ and distribute uniformly along the reference direction on the feeding patch.

Basically, five Q factors are compared. $Q_{po}$, $Q_Z$, and $Q_{FBW}$ are respectively calculated with (33), (34) and (35). $Q_F$, $Q_{van}$ are calculated in the way similar to $Q_{po}$, but with the reactive energies replaced by $W_F$ and $W_{van}$, respectively. Unless specified differently, the VSWR used to determine the bandwidth is 1.5 and the corresponding Q factor is denoted by $Q_{FBW-1.5}$. $Q_F$ is calculated with the origin located at the center of the structure.

A fractional discrepancy between the Q factors obtained with the new formulation and that with Vandenbosch formulation is defined as,

$$\Delta Q = \left|Q_{po} - Q_{van}\right| / Q_{van} \times 100\% \quad (61)$$

A mathematical uniform surface current ring is analyzed firstly. Similar to that in [52], the current is not associated with a real metal plate, but is a purely feeding source distributed over a circular ring with radius of 15mm and width of 0.5mm. The phase of the current is $e^{-j2\varphi}$, which varies linearly along the circle. The calculated energies and Q factors are shown in Fig.2(a). It can be seem that all the reactive electric energies ($WE_{van}$, $WE_F$) and reactive magnetic energies ($WM_{van}$, $WM_F$) calculated with Vandenbosch formulation and formulae of $W_F$ are negative near 29GHz, where the reactive energies ($W_{\rho\varphi}$, $W_{AJ}$) calculated with source-potential formulation are all positive.

The Q factors are shown in Fig.2(b). $Q_F$ is calculated with the origin located at the center of the ring, and almost coincides with $Q_{van}$ in this example. Owing to the negative energies, $Q_{van}$ and $Q_F$ are negative near 29GHz. It can be noted that there is a jump in $Q_{FBW-1.5}$ at 25.6GHz, which is caused by a neighboring local resonance. $Q_{po}$ is always positive.

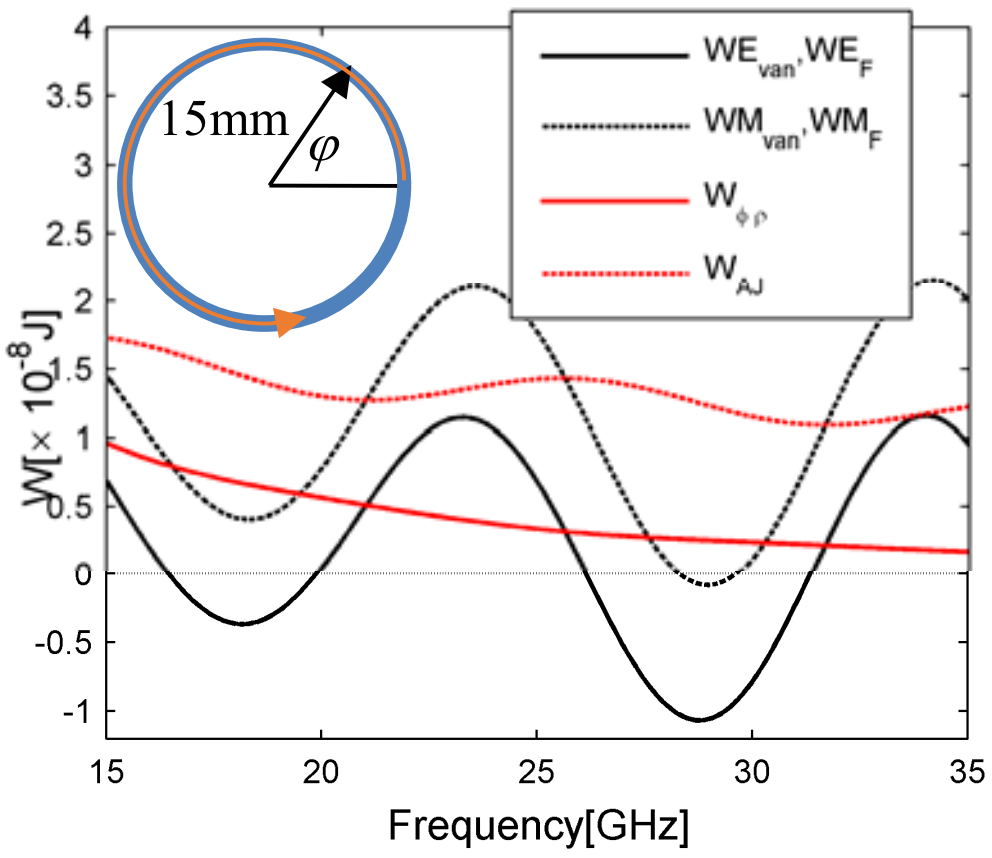

(a)

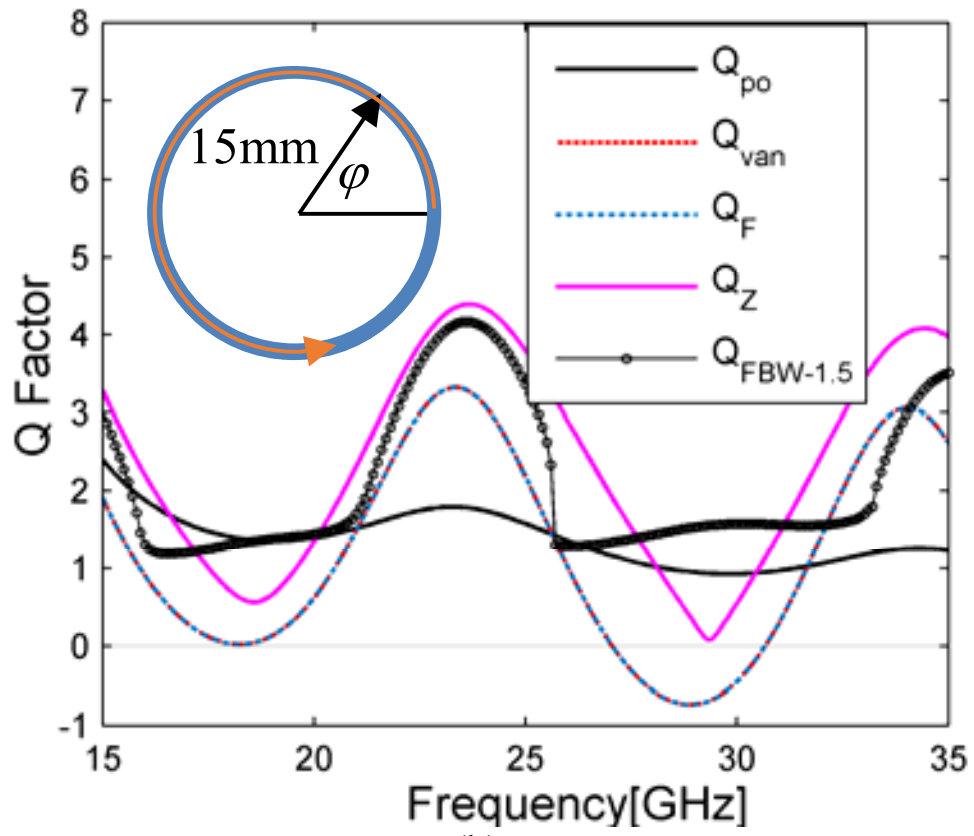

(b)

Fig.2. Mathematical surface current over a ring. (a) Energies. (b) Q Factors.

Next, a plate dipole is analyzed. It consists of two PEC plates with size of $500\text{mm}\times 2\text{mm}$. A feeding patch with size of $2\text{mm}\times 2\text{mm}$ is placed between the two plates. The surface current on the antenna is calculated by solving the corresponding EFIF with Galerkin method. The results of the input resistance $R$ and reactance $X$ are shown in Fig.3(a), and the calculated Q factors are plotted in Fig.3(b). It can be seen that all the five Q factors are roughly close to each other in the examined frequency band. The relative discrepancy $\Delta Q$ is less than 16% in this example. Although not plotted in the figure, it has been checked that the effect of the origin is quite small in this case due to the symmetrical property of the fields. It has also been checked that when the VSWR is set to be 2.0, 1.5, and 1.05, the corresponding Q factors, denoted respectively by $Q_{FBW-2.0}$, $Q_{FBW-1.5}$, and $Q_{FBW-1.05}$, are all close to $Q_Z$ (only $Q_{FBW-1.5}$ is plotted.)

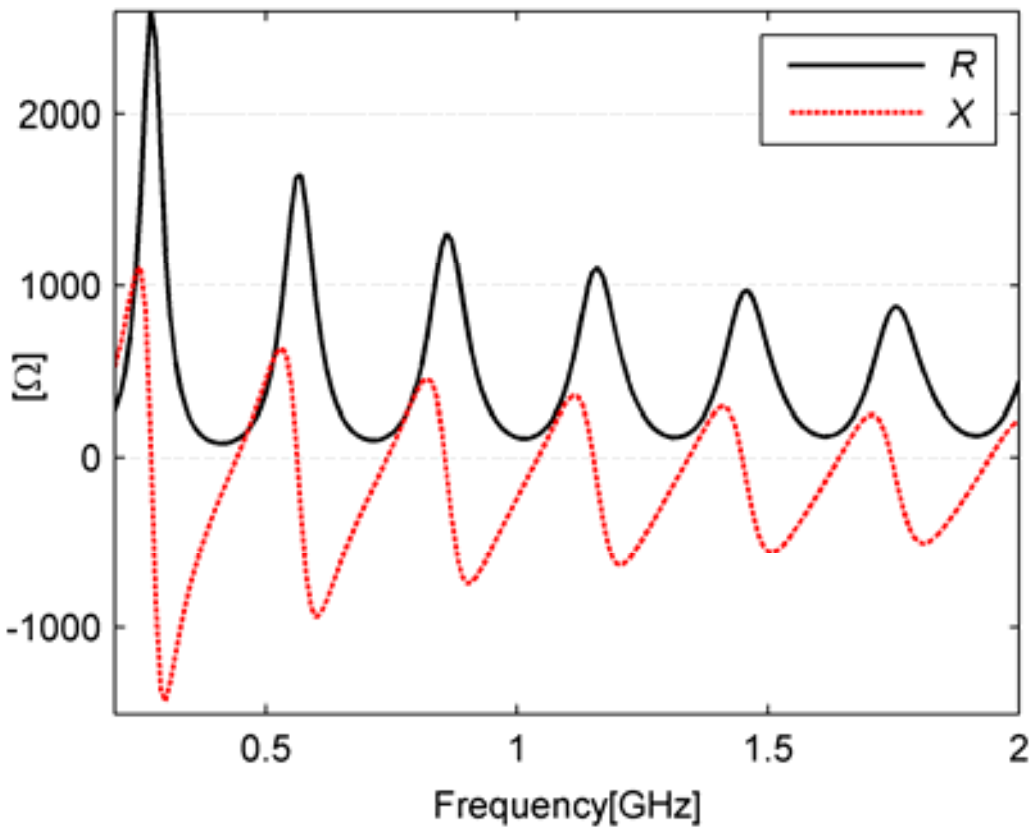

(a)

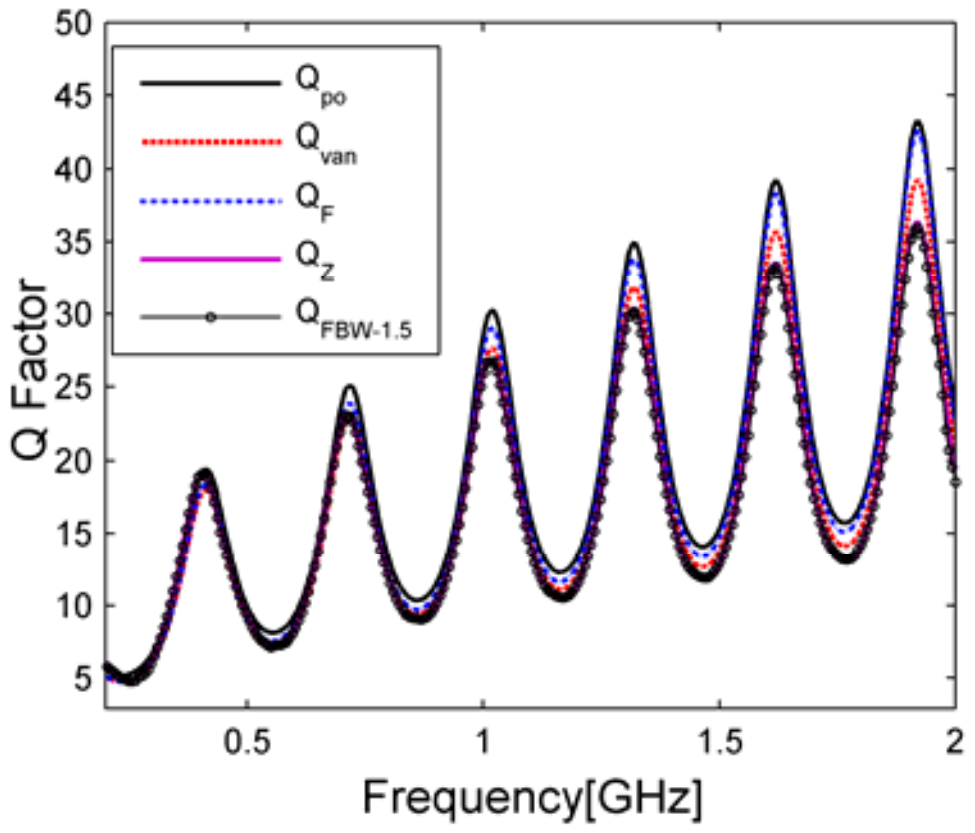

(b)

Fig. 3. A PEC plate dipole. (a) Input resistance and reactance. (b) Q factors.

The third example is a square loop antenna with edge length of 30mm, as shown in Fig. 4(a), excluding the PEC ground plate. The width of the PEC strip is 0.5mm. A $0.5\text{mm}\times0.5\text{mm}$ feeding patch is put at the center of one segment of the square. The results of Q factors are plotted in Fig. 4(b). The five Q factors reveal similar behavior but with larger discrepancies than that in the second example.

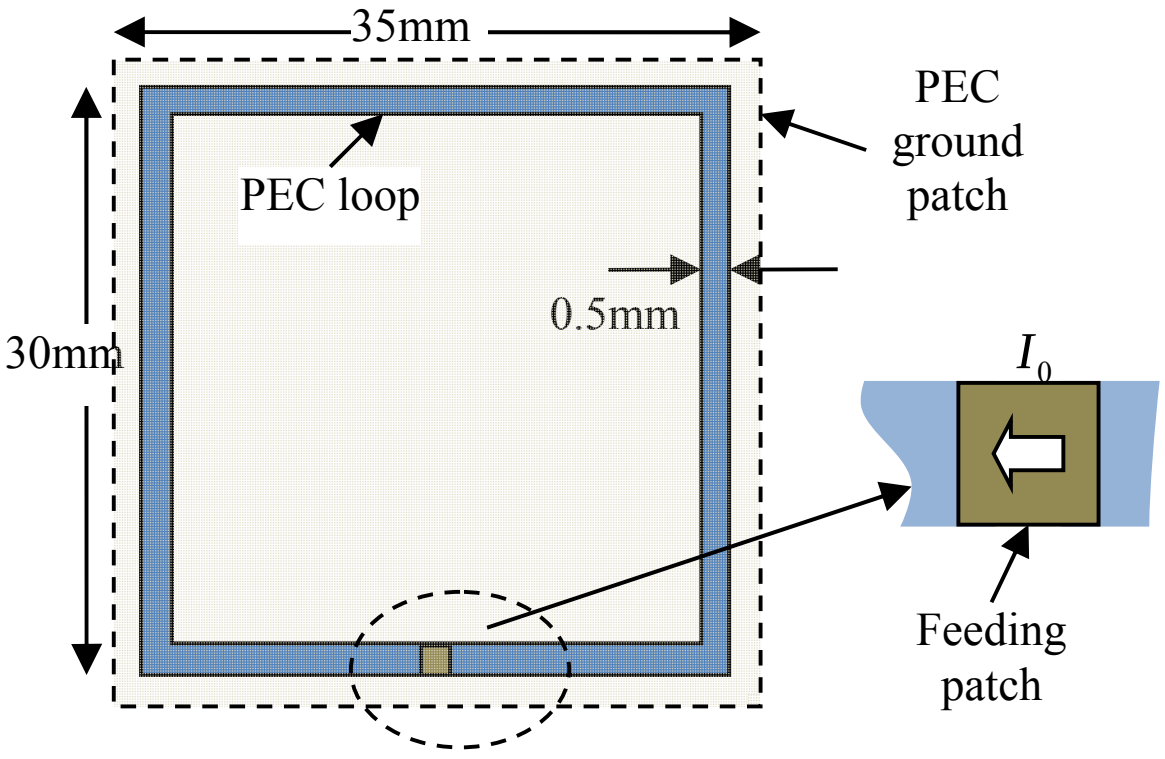

(a)

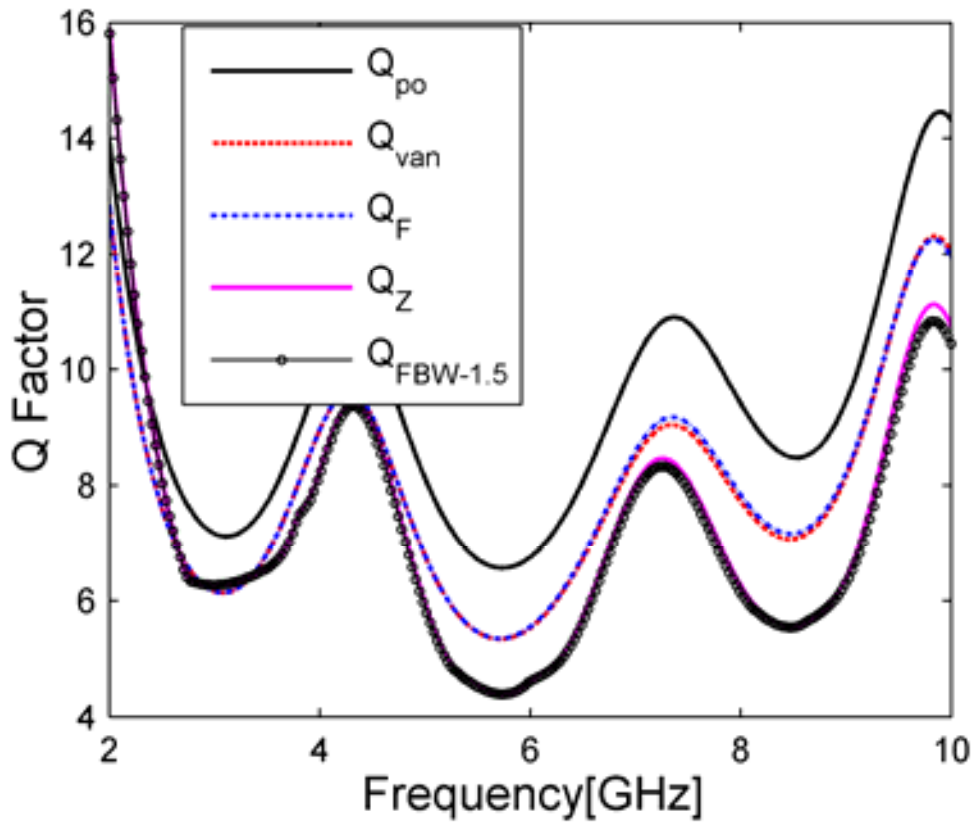

(b)

Fig. 4. A PEC plate loop. (a) Square loop structure. (b) Q factors.

In the fourth example, a PEC plate with size of $35\text{mm}\times35\text{mm}$ is placed under the loop antenna, as shown with dot lines in Fig. 4(a). When the ground plane is 2mm away, the calculated Q factors are plotted in Fig.5(a). It can be seen that $Q_{po}$, $Q_{van}$ and $Q_F$ are very close. $Q_Z$ and $Q_{FBW}$ agree well with each other, and also roughly agree with the other three Q factors. However, $Q_{FBW}$ has spurs near the natural resonances, where the input impedance varies sharply, as can be seen from Fig. 5(b). With the increase of *h*, the distance between the loop and ground plate, the Q factors gradually decease and approach to those of the loop without ground, as shown in Fig. 5(c). The relative discrepancies are shown in Fig. 5(d). The discrepancy tends to become smaller when the ground plate approaches the loop.

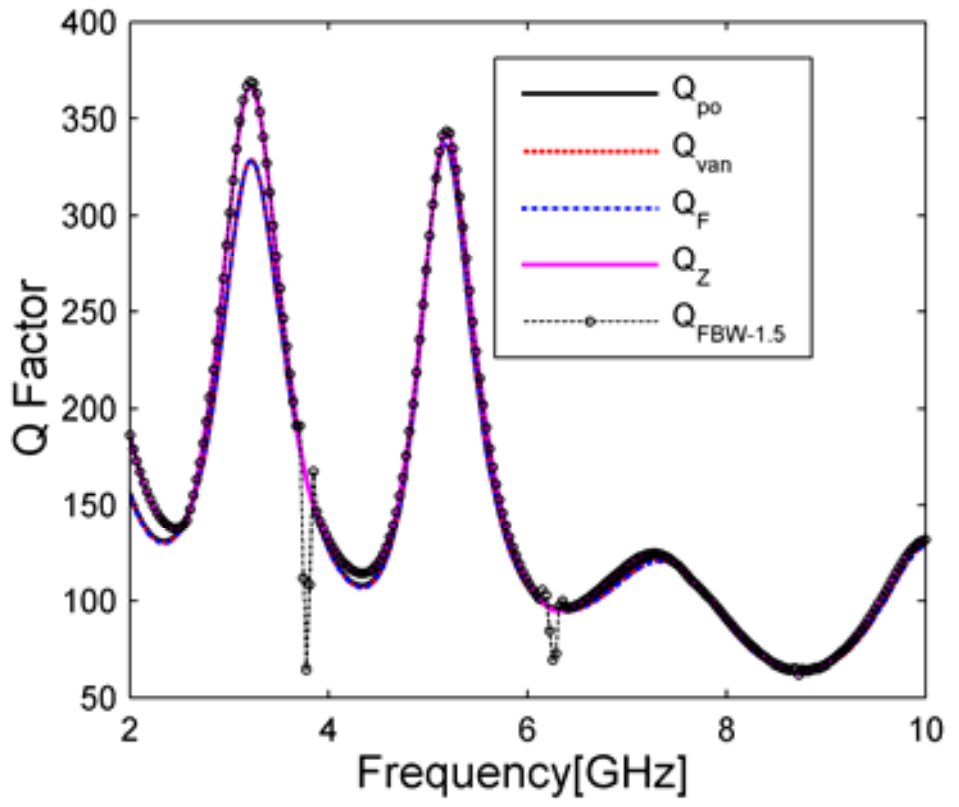

(a)

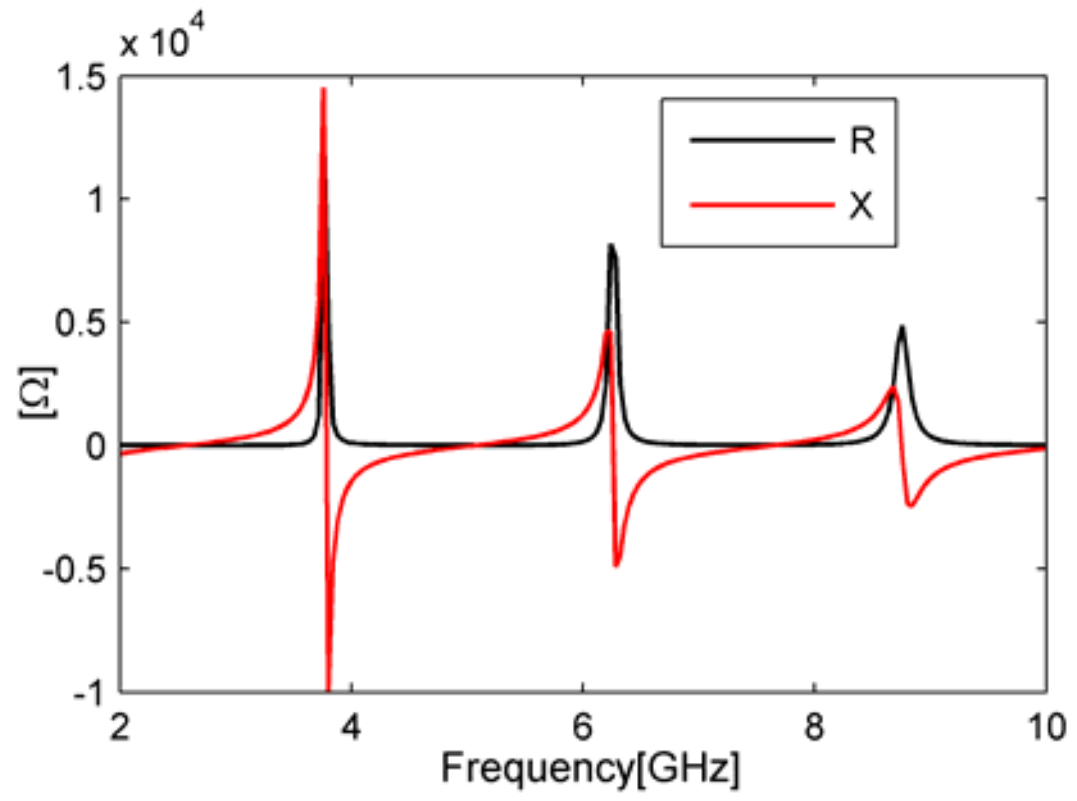

(b)

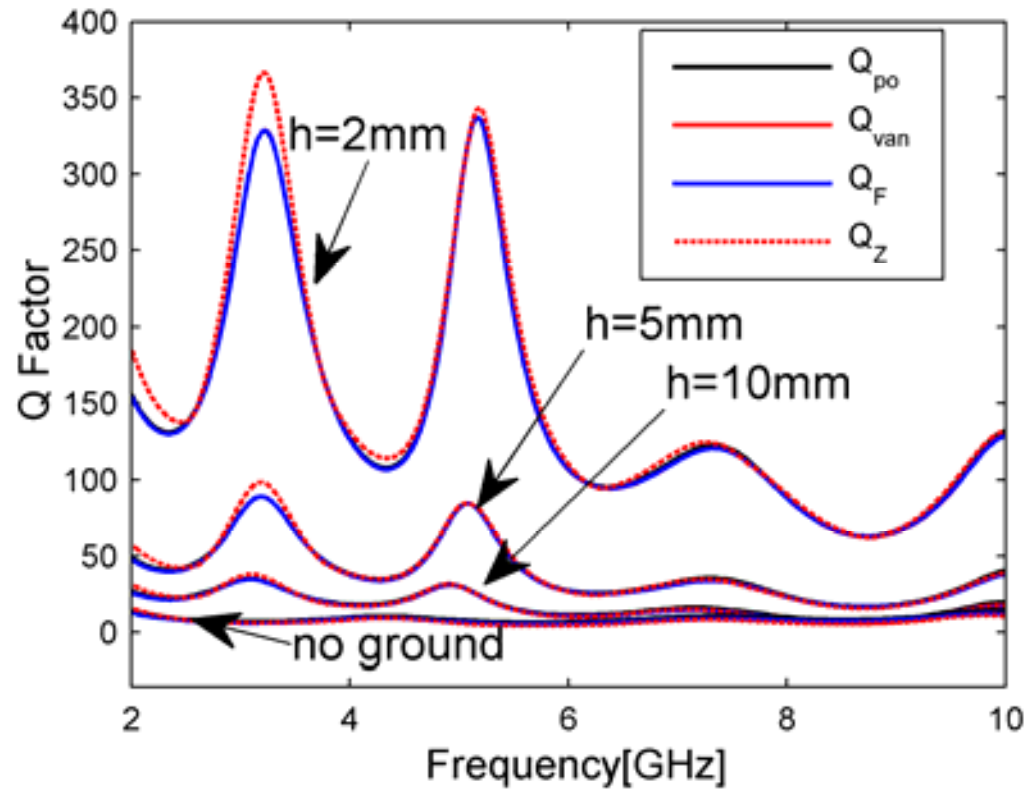

(c)

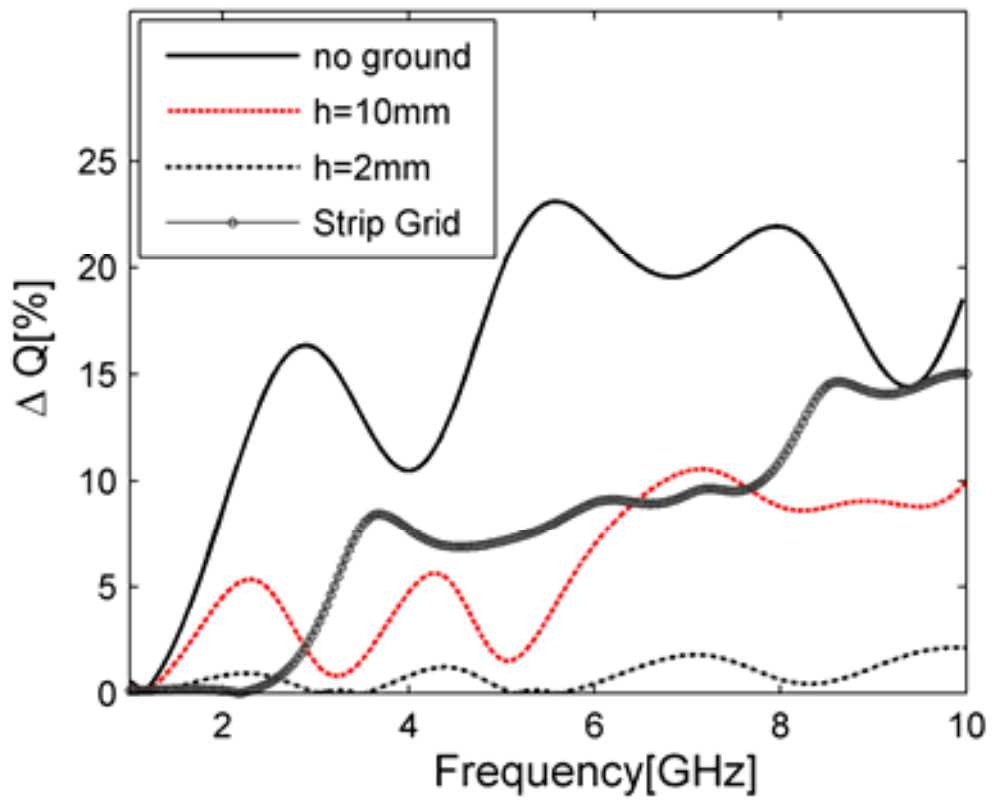

(d)

Fig. 5. A PEC square loop antenna on a PEC square plate. (a) Q factors @$h$=2mm. (b) Input resistance and reactance @$h$=2mm. (c) Q factors for different $h$. (d) Relative discrepancy of Q factors.

The fifth example is a PEC strip grid antenna consists of uniform strips with width of 0.8mm. The antenna is symmetrical, but the feeding point is usually offset to get optimal matching performance. The calculated Q factors are shown in Fig.6, from which it can be seen that $Q_F$ and $Q_{van}$ almost coincide, and $Q_{po}$ is close to them, with discrepancy illustrated in Fig. 6(b). Due to the effect of local resonances, $Q_{\mathrm{FBW-1.5}}$ does not agree with $Q_Z$ at frequency range of about 8~9GHz. The influence of the choice of VSWR on $Q_{\mathrm{FBW}}$ is illustrated in Fig.6(c), where $Q_{\mathrm{FBW-1.05}}$ almost coincides with $Q_Z$ , but $Q_{\mathrm{FBW-1.5}}$ and $Q_{\mathrm{FBW-2.0}}$ may have jumps in them. Although larger VSWR causes larger bandwidth, the corresponding $Q_{\mathrm{FBW}}$ is not necessary to be smaller because its definition includes the VSWR ( $s$ ), as can be seen from Fig.6(c).

The effect of the choice of origin is also illustrated in Fig.6(c). The origin is chosen to locate uniformly at 20 points on the yellow dot-lined circle shown in Fig.6(a). The resulted $Q_F$ s are shown with grey lines in Fig.6(c). Although the effect is bounded, it is not always neglectable.

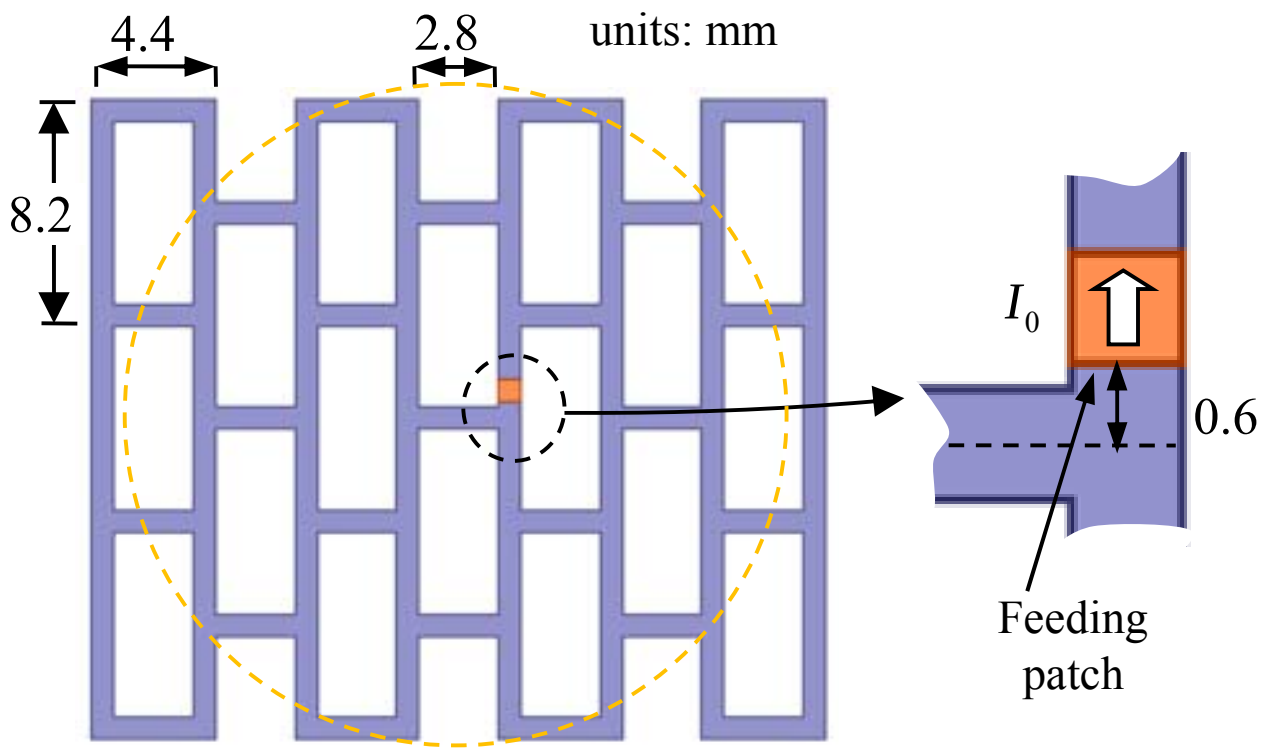

(a)

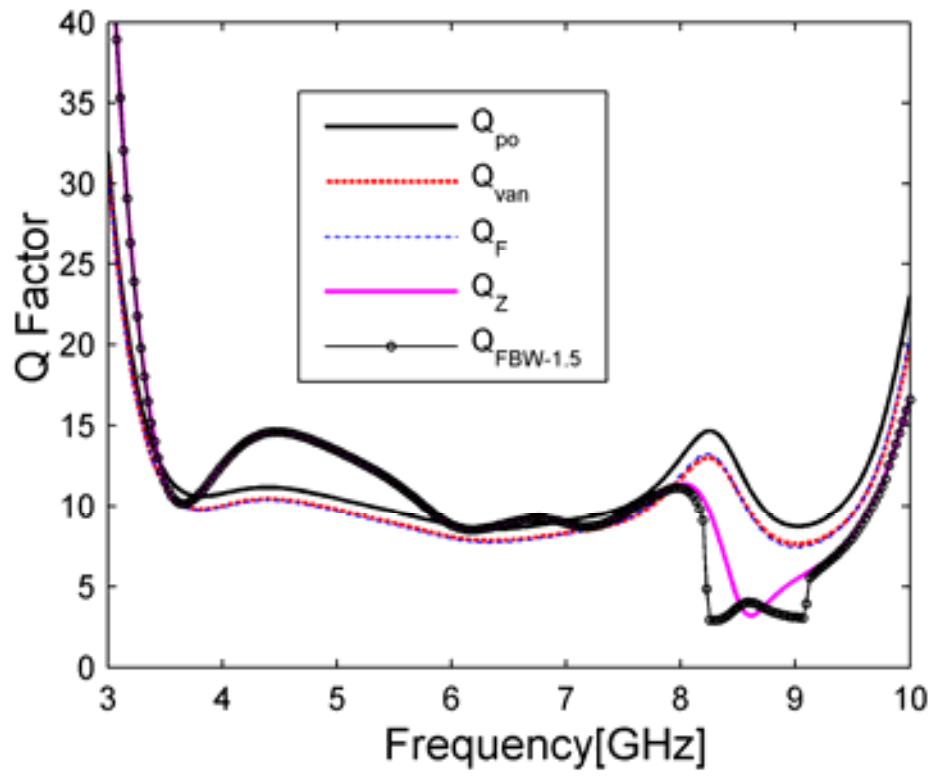

(b)

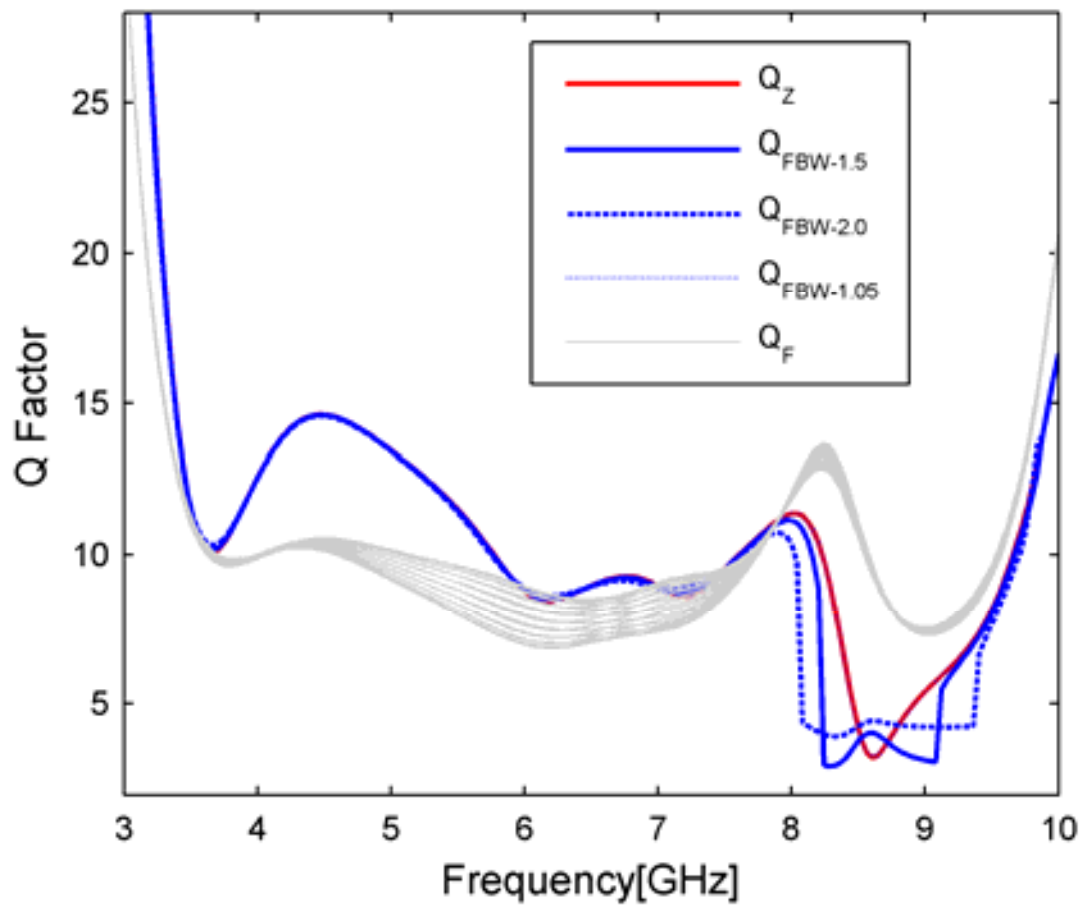

(c)

Fig. 6. A PEC strip grid antenna. (a) Structure and parameters (Units: mm). (b) Q factors. (c) Effect of VSWR and origin.

The last example is a Vivaldi antenna with structure shown in Fig.7(a). The opening rate is $R_a = 0.0458/\mathrm{mm}$ [53]. The calculated input resistance and reactance are shown in Fig.7(b). Since its resistance is near 50ohm in a very wide frequency range, it is easy to realize a wide band antenna with further optimization. The Q factors are shown in Fig.7(c). Similar to the results of the strip grid antenna, $Q_{po}$, $Q_{van}$ and $Q_F$ (with origin at the center of the feeding patch) are close to each other with a relative discrepancy at most 18%. However, $Q_{FBW}$ and $Q_Z$ are obviously away from them. The effect of the choice of VSWR is illustrated in Fig.7(d). Since the Vivaldi antenna is basically a wideband antenna, the variation of $Q_{FBW-2.0}$ is not quite smooth because many local resonances may fall into the pass band at a specified tuned frequency. By changing the origin to 20 points located on the yellow dot-lined circle in Fig. 7(a), the resultant $Q_F$ s are shown with the grey line in Fig. 7(d). Since the variation range of $Q_F$ is proportional to the offset distance of the origin,

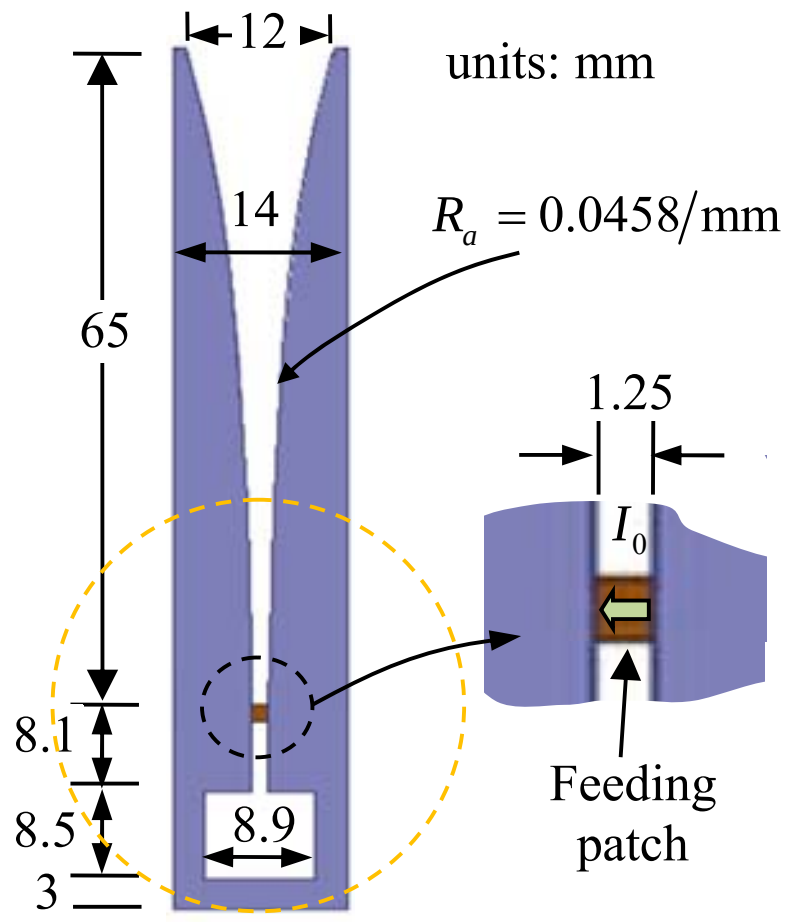

(a)

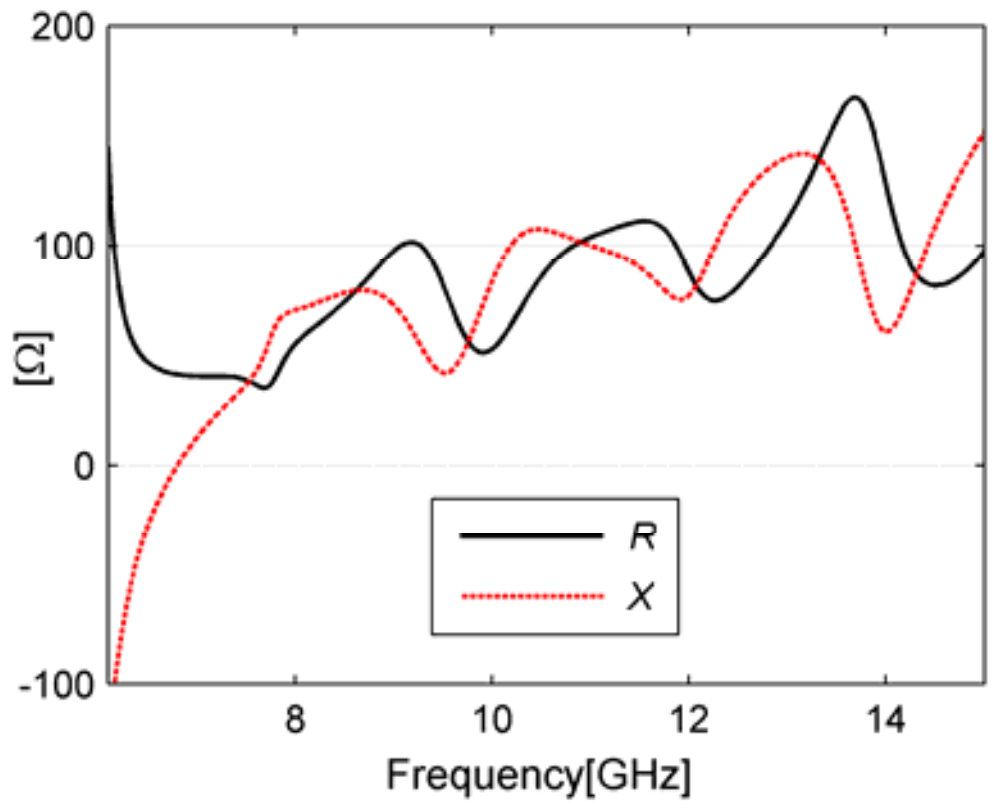

(b)

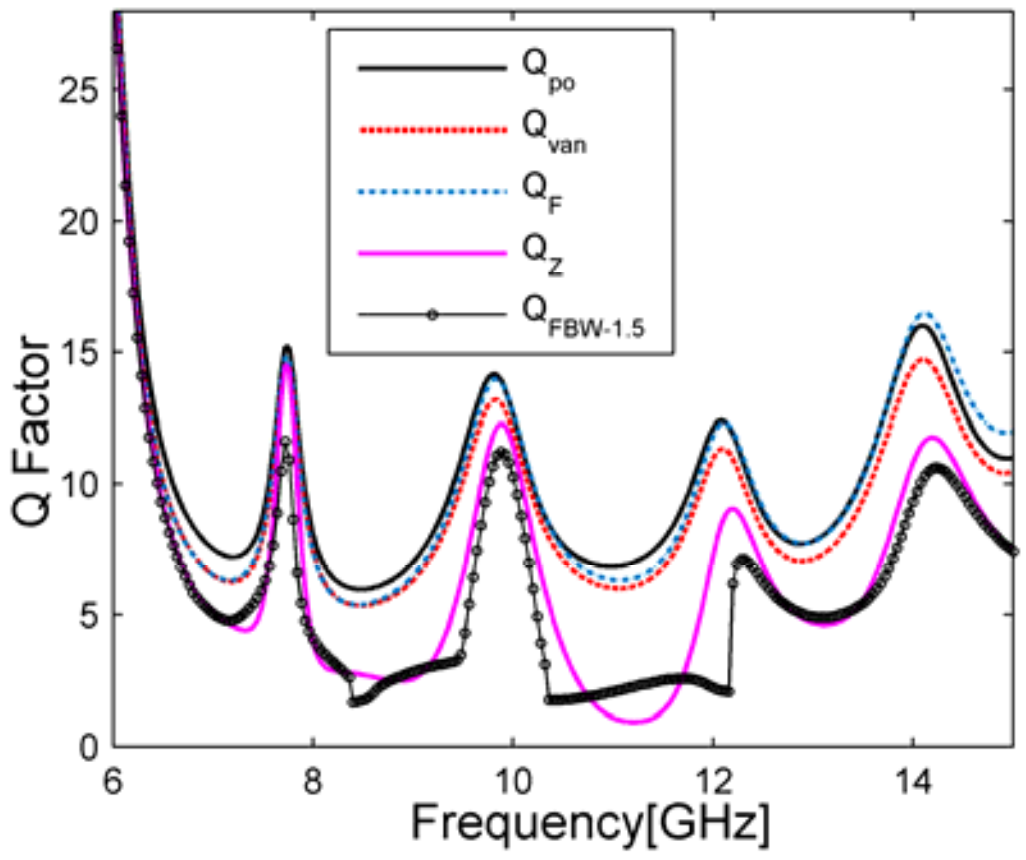

(c)

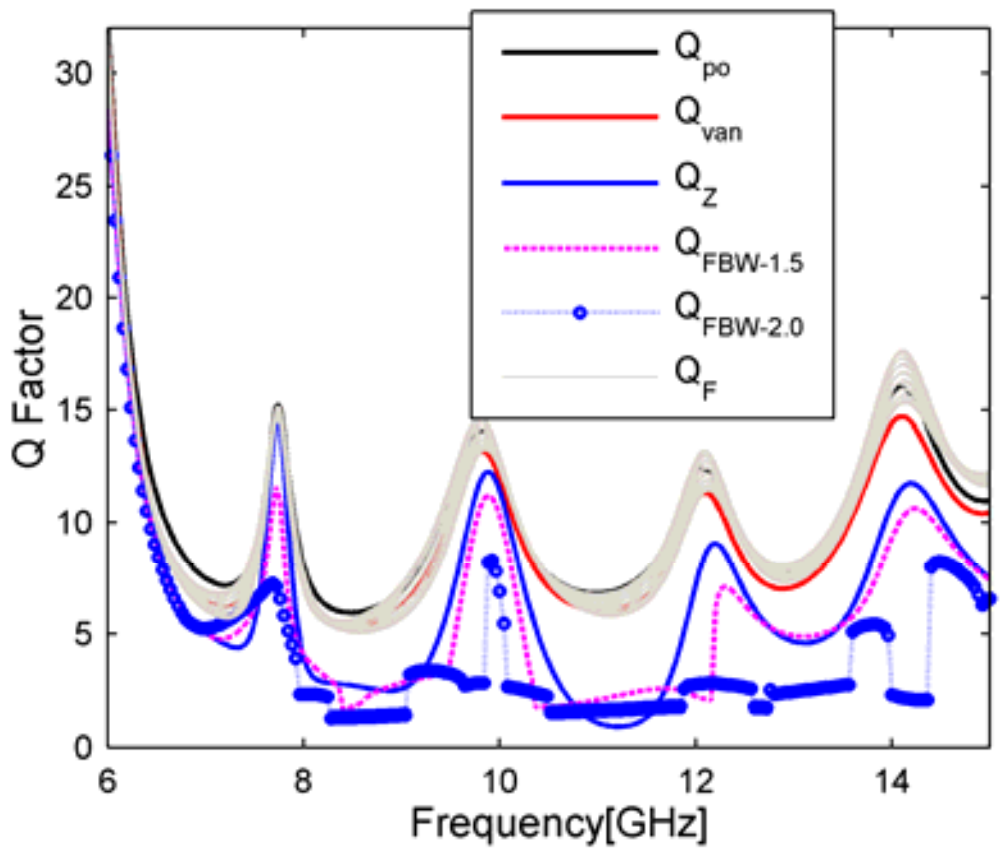

Fig. 7. Vivaldi antenna. (a) Structure and parameters. (b) Input resistance and reactance. (c) Q factors. (d) Effect of VSWR and origin.

All these examples show that $Q_{po}$ computed using the new formulation are always close to $Q_{van}$ with relatively small discrepancies. This is not strange since the main body of the expressions for the reactive energies of the two formulations is identical. $Q_F$ usually depends on the choice of origin, and the variation range is proportional to the offset distance. For antennas with symmetrical radiation fields, the effect may be very small.

The numerical results also show that $Q_Z$ and $Q_{FBW}$ calculated using Yaghjian-Best formulation are also close to $Q_{po}$ and $Q_{van}$ for narrow band radiators, such as the dipoles, but may differ largely from them for wide band radiators, as shown in the last two examples.

## VII. Conclusion

A new formulation is proposed to calculate the stored reactive energy of a radiator based on sources and potentials for harmonic fields. Similar to the conventional definitions for the electromagnetic reactive energies, the new definition also lacks rigorous mathematical proof. However, some arguments are provided to support the new formulation. Particularly, the case study of Hertzian dipole has verified that the stored reactive energies evaluated by the new formulation exactly agree with the classical results obtained based on Chu's circuit model. On the other hand, by analyzing the energy balance relationship in a radiation problem, the new formulation can provide an insight into the composition of the energies radiated by an antenna. The expressions for the reactive energies are quite simple and easy to implement. Q factors can be evaluated at a single frequency point. At some situations in which the conventional methods may take negative reactive energies while the new formulation still provides positive results.

This work focuses on the reactive electromagnetic energies of radiators in free space or radiation problems that can be handled with equivalent models in free space. Further investigations are needed on systems consisting of isotropic, lossy or time varying media, as well as the mathematical aspect of the new formulation.

## Appendix A

In order to show that the real part of $\oint_{S_\infty}\left(\vec{H}\times\vec{A}^*\right)\cdot d\vec{S}$ in (29) approaches zero, we have to use the asymptotic behavior of the scalar Green's function in 3D free space. Since in this case the integration is performed for time harmonic fields, we can write

$$\oint_{S_\infty}\left(\vec{H}\times\vec{A}^*\right)\cdot d\vec{S} = \mu\oint_{S_\infty}\int_{V_a}\int_{V_a}\left(\nabla g_1\cdot\vec{J}_2^*\right)\left(\vec{J}_1\cdot\hat{r}g_2^*\right)d\vec{r}_2 d\vec{r}_1 dS - \mu\oint_{S_\infty}\int_{V_a}\int_{V_a}\left(\vec{J}_1\cdot\vec{J}_2^*\right)\left(g_2^*\nabla g_1\cdot\hat{r}\right)d\vec{r}_2 d\vec{r}_1 dS \quad (62)$$

where $g_{1,2} = g\left(\vec{r},\vec{r}_{1,2}\right)$, $\vec{J}_{1,2} = \vec{J}\left(\vec{r}_{1,2}\right)$. Furthermore, we have the asymptotic expressions

$$\lim_{r\to\infty} g_{1,2} = \frac{e^{-jkr}}{4\pi r}e^{-j\vec{k}\cdot\vec{r}_{1,2}} \quad (63)$$

$$\lim_{r\to\infty}\nabla g_{1,2} \sim \frac{-jke^{-jkr}}{4\pi r}e^{-j\vec{k}\cdot\vec{r}_{1,2}}\hat{r} \approx -jkg_{1,2}\hat{r} \quad (64)$$

Making use of the relationship that

$$\lim_{r\to\infty}\left(\nabla g_2^*\cdot\vec{J}_1\right)\left(\vec{J}_2^*\cdot\nabla g_1\right) \approx \lim_{r\to\infty}\left(\nabla g_1\cdot\vec{J}_1\right)\left(\vec{J}_2^*\cdot\nabla g_2^*\right) \quad (65)$$

and

$$\int_{V_a}\nabla_{1,2}\cdot\left(g_{1,2}\vec{J}_{1,2}\right)d\vec{r}_{1,2} = 0 \quad (66)$$

It can be deduced that

$$\oint_{S_\infty}\int_{V_a}\int_{V_a}\left(\nabla g_1\cdot\vec{J}_2^*\right)\left(\vec{J}_1\cdot\hat{r}g_2^*\right)d\vec{r}_2 d\vec{r}_1 dS=\frac{1}{jk}\oint_{S_\infty}\int_{V_a}\int_{V_a}\left(\nabla_1 g_1\cdot\vec{J}_1\right)\left(\vec{J}_2^*\cdot\nabla_2 g_2^*\right)d\vec{r}_2 d\vec{r}_1 dS=\frac{1}{jk}\int_{V_a}\int_{V_a}\left(\nabla_1\cdot\vec{J}_1\right)\left(\nabla_2\cdot\vec{J}_2^*\right)\oint_{S_\infty}g_1 g_2^* dS d\vec{r}_2 d\vec{r}_1 \quad (67)$$

$$-\oint_{S_\infty}\int_{V_a}\int_{V_a}\left(\vec{J}_1\cdot\vec{J}_2^*\right)\left(g_2^*\nabla g_1\cdot\hat{r}\right)d\vec{r}_2 d\vec{r}_1 dS\approx jk\int_{V_a}\int_{V_a}\left(\vec{J}_1\cdot\vec{J}_2^*\right)\oint_{S_\infty}g_2^* g_1 dS d\vec{r}_2 d\vec{r}_1 \quad (68)$$

Using (63) and performing the integration directly gives

$$\oint_{S_\infty}g_2^* g_1 dS=-\frac{\sin k\left|\vec{r}_1-\vec{r}_2\right|}{4\pi k\left|\vec{r}_2-\vec{r}_2\right|} \quad (69)$$

Hence we have

$$\oint_{S_\infty}\left(\vec{H}\times\vec{A}^*\right)\cdot d\vec{S}=-\frac{j\mu}{4\pi k}\int_{V_a}\int_{V_a}\left\{\left[k^2\vec{J}_1\cdot\vec{J}_2^*-\left(\nabla_1\cdot\vec{J}_1\right)\left(\nabla_2\cdot\vec{J}_2^*\right)\right]\times\frac{\sin k\left|\vec{r}_1-\vec{r}_2\right|}{k\left|\vec{r}_2-\vec{r}_2\right|}\right\}d\vec{r}_2 d\vec{r}_1 \quad (70)$$

where it is understood that the two currents in the two-fold integral are the same one but at different locations in the same region $V_a$. By exchanging $\vec{r}_1$ and $\vec{r}_2$, together with the order of the inner-fold and the outer-fold integration, it is straightforward to check that

$$\oint_{S_\infty}\left(\vec{H}\times\vec{A}^*\right)\cdot d\vec{S}=-\oint_{S_\infty}\left(\vec{H}\times\vec{A}^*\right)^*\cdot d\vec{S} \quad (71)$$

which leads to $\mathrm{Re}\oint_{S_\infty}\left(\vec{H}\times\vec{A}^*\right)\cdot d\vec{S}=0$.

## Appendix B

Inserting the scalar potential into (14) and applying the current continuity equation yields

$$\tilde{W}_E=\mathrm{Re}\frac{1}{4}\int_{V_\Sigma}\rho_\Sigma\left(\vec{r}_1\right)\phi^*\left(\vec{r}_1\right)d\vec{r}_1=\mathrm{Re}\frac{1}{4}\int_{V_\Sigma}\rho_\Sigma\left(\vec{r}_1\right)\frac{1}{\varepsilon}\int_{V_\Sigma}\rho_\Sigma^*\left(\vec{r}_2\right)g^*\left(\vec{r}_1,\vec{r}_2\right)d\vec{r}_2 d\vec{r}_1=\frac{1}{4\omega^2\varepsilon}\mathrm{Re}\int_{V_\Sigma}\int_{V_\Sigma}\nabla_1\cdot\vec{J}_\Sigma\left(\vec{r}_1\right)\nabla_2\cdot\vec{J}_\Sigma^*\left(\vec{r}_2\right)g^*\left(\vec{r}_1,\vec{r}_2\right)d\vec{r}_2 d\vec{r}_1$$

$$=\frac{\mu}{16\pi k^2}\mathrm{Re}\int_{V_\Sigma}\int_{V_\Sigma}\nabla_1\cdot\vec{J}_\Sigma\left(\vec{r}_1\right)\nabla_2\cdot\vec{J}_\Sigma^*\left(\vec{r}_2\right)\frac{\cos kR}{R}d\vec{r}_2 d\vec{r}_1+\frac{\mu}{16\pi k^2}\mathrm{Re}\int_{V_\Sigma}\int_{V_\Sigma}\nabla_1\cdot\vec{J}_\Sigma\left(\vec{r}_1\right)\nabla_2\cdot\vec{J}_\Sigma^*\left(\vec{r}_2\right)\frac{j\sin kR}{R}d\vec{r}_2 d\vec{r}_1 \quad (72)$$

With the same analogy to prove (71), it is straightforward to check that the first two-fold integral at the RHS of the last equation is pure real, while the second one is pure imaginary. Hence (54) is proved.

Inserting the vector potential into (16) and applying the current continuity equation yields

$$\begin{aligned}\tilde{W}_M&=\mathrm{Re}\frac{1}{4}\int_{V_\Sigma}\vec{J}_\Sigma\left(\vec{r}_1\right)\cdot\vec{A}^*\left(\vec{r}_1\right)d\vec{r}_1=\mathrm{Re}\frac{\mu}{4}\int_{V_\Sigma}\vec{J}_\Sigma\left(\vec{r}_1\right)\cdot\int_{V_\Sigma}\vec{J}_\Sigma^*\left(\vec{r}_2\right)g^*\left(\vec{r}_1,\vec{r}_2\right)d\vec{r}_2 d\vec{r}_1\\&=\frac{\mu}{16\pi}\mathrm{Re}\int_{V_\Sigma}\vec{J}_\Sigma\left(\vec{r}_1\right)\cdot\int_{V_\Sigma}\vec{J}_\Sigma^*\left(\vec{r}_2\right)\frac{\cos kR}{R}d\vec{r}_2 d\vec{r}_1+\frac{\mu}{16\pi}\mathrm{Re}\int_{V_\Sigma}\vec{J}_\Sigma\left(\vec{r}_1\right)\cdot\int_{V_\Sigma}\vec{J}_\Sigma^*\left(\vec{r}_2\right)\frac{j\sin kR}{R}d\vec{r}_2 d\vec{r}_1\end{aligned} \quad (73)$$

(55) can be obtained in the same way to derive (54).

It can also be verified that, if adopting the identity (A.1) in [25], (54) and (55) can be derived respectively from the direct integration of the energy densities (17) and (19) over the entire space.